\documentclass{article}
\usepackage{float}
\usepackage{graphicx}

\usepackage{caption}
\usepackage{subcaption}
\usepackage{amssymb}
\usepackage{amsmath}
\usepackage{amsthm}
\usepackage{enumitem}
\usepackage{mdwlist}
\usepackage{bbm}
\usepackage{array}
\usepackage{parskip}
\usepackage[top=1.5in,bottom=1.5in,right=1.75in, left=1.75in]{geometry}
\usepackage{todonotes}
\usepackage[authoryear,comma,longnamesfirst,sectionbib]{natbib} 
\usepackage{url} 

\makeatletter
\newtheoremstyle{indented}
  {8pt}
  {8pt}
  {\setlength{\leftskip}{2.5em}\addtolength{\@totalleftmargin}{2.5em}}
  {}
  {\bfseries}
  {.}
  {.5em}
  {}
\makeatother

\makeatletter
\newtheoremstyle{proofstyle}
  {8pt}
  {8pt}
  {\setlength{\leftskip}{2.5em}\addtolength{\@totalleftmargin}{2.5em}}
  {}
  {\itshape}
  {.}
  {.5em}
  {}
\makeatother

\theoremstyle{indented}

\theoremstyle{proofstyle}

\def \cov {\text{Cov}}
\def \cond {\,|\,}

\newcommand{\ind}[1]{\mathbbm{1}_{#1}}

\DeclareMathOperator*{\argmin}{arg\,min}

\newcommand{\expec}{\mathbb{E}}

\newcommand{\eps}{\epsilon}
\newcommand{\lam}{\lambda}

\newcommand{\bs}[1]{\boldsymbol{#1}}

\title{Functional Data Analysis of Aging Curves in Sports}
\author{Alexander Wakim and Jimmy Jin}

\begin{document}
\maketitle

\tableofcontents

\newpage

\section{Introduction}
It is well known that athletic and physical condition is affected by age. Plotting an individual athlete's performance against age creates a graph commonly called the player's \textit{aging curve}. Despite the obvious interest to coaches and managers, the analysis of aging curves so far has used fairly rudimentary techniques (e.g. multiple regression analysis with interpolating splines) that limit the scope of the conclusions (\citealt{Fair2008}). In this paper, we introduce functional data analysis (FDA) to the study of aging curves in sports and argue that it is both more general and more flexible compared to the methods that have previously been used. First, we review traditional FDA techniques and talk about a specialized method called PACE which is particularly useful for sports analysis. Second, we highlight the advantages of functional data analysis in sports. Finally, we illustrate the rich analysis that is possible when taking an FDA approach by analyzing data for NBA and MLB players. 

In the analysis of MLB data, we perform functional hypothesis testing to show differences in aging curves between potential power hitters and potential non-power hitters. We also illustrate a tool that allows us to discern the major modes of variation in the ensemble of aging curves of MLB players.

The analysis of aging curves in NBA players illustrates the use of the PACE method. Using a functional framework, we show that there are three distinct aging patterns in NBA players. Looking at the players that fall into each of the three patterns, we see that there are differences in scoring ability among the players that fall in different aging patterns. We also show that aging pattern is independent of position. This demonstrates how the rich analysis of FDA can lead to conclusions that differ from popular opinion and analyses that use more rudimentary methods.

\section{Functional Data Analysis Methodology}

\subsection{Traditional Functional Data Analysis}
FDA begins by assuming that performance across age is a smooth, and hence continuous, function. We call this function a player's \textit{aging curve}. However, in practice, we observe discrete performance metrics at certain ages, such as Win Shares (WS) for the NBA or weighted on-base average (wOBA) for the MLB, that come from this function.  Thus, we must fit the smooth aging curve $f_i$ from which the observed data comes from. More formally, traditional FDA assumes that for each datum $i = 1, \ldots, N$ we observe $n$ values $y_{i1}, \ldots, y_{in}$ measured at ages $t_{1}, \ldots, t_{n}$. We express this as
\begin{equation}
y_{ij} = f_i(t_{ij}) + \eps_{ij}
\end{equation}
with error term $\eps_{ij}$ contributing possible roughness to the observed data. In section 2.2 we will discuss a specialized method called PACE that can handle situations where the number of observed measurements $n$ can differ across players.  

In multivariate data analysis, one analyzes and summarizes vectors. Analogously, in FDA one analyzes and summarizes fitted functions. In our case, our fitted functions are aging curves. In order to perform analysis on the aging curves $f_i(t)$, we must fit them from the observed performance metrics $y_{i1}, \ldots, y_{in}$. To do this, we use basis functions to get arbitrarily good approximations to the curves $f_i(t)$. Basis functions are an orthogonal set of functions $\phi_1(t), \phi_2(t), \phi_3(t), \ldots$ which have the property that for any function $f(t)$ we have
\begin{equation} \label{basisprop}
f(t) = \sum_{j=1}^{\infty} \beta_j \phi_j(t)
\end{equation}
for some constants $\beta_j$. The analysis in this paper will use a basis called the B-spline basis which consists of functions that are piecewise polynomial of degree $k$  defined over a range $[\tau_0, \tau_{m+1}]$ where $m \geq k + 2$ (\citealt[p.~87-142]{deBoor2001}). Here $\tau_1, \tau_2, \ldots, \tau_m$ are the \textit{interior knots} where the pieces of the piecewise polynomial meet and $\tau_0$ and $\tau_{m+1}$ are the \textit{end points}. So, we re-write equation \eqref{basisprop} as
\begin{equation}
f(t) = \sum_{j=1}^{\infty} \beta_j B_j(t, \tau)
\end{equation}
where $B_j(t, \tau)$ indicates the values at point $t$ of the B-spline basis function defined by knot sequence $\tau$. In practice, we have the property that for any smooth function $f(t)$ there exists a spline function $f^*(t)$ on $[t_0, t_{m+1}]$ such that 
\begin{equation}
f(t) \approx f^*(t) = \sum_{j=1}^J \beta_j B_j(t,\tau)
\end{equation}  
for sufficiently large J (\citealt[p.~149]{deBoor2001}). Typically, we set $J=k+m$ (\citealt[p.~49]{RamsaySilverman2005}). Note that any spline function of degree $k$ on a given set of knots can be uniquely expressed as a finite linear combination of B-splines whose coefficients $\bs{\beta} = (\beta_1, \ldots, \beta_J)'$ uniquely describe it (\citealt[p.~101]{deBoor2001}). In summary, using a B-spline basis allows us to uniquely approximate any smooth function.

The coefficients $\bs{\beta}$ uniquely describe the function and so estimating them is of interest. In estimating the fitted function, we want a good fit for the data that does not over-fit. That is, the goal is often to find a function that captures important patterns in the data without over-compensating for noise. So, to estimate the fitted function we add a \textit{roughness penalty} to the least-squares criterion 
\begin{equation} \label{ls-penalized}
\hat{\bs{\beta}}_i = \argmin_{\beta_1, \ldots, \beta_J} \sum_{k=1}^n \Big( y_{ik} - \sum_{j=1}^J \beta_j B_j(t,\tau) \Big)^2 + \lambda \int \left\{ \Big( \sum_{j=1}^J \beta_j B_j(t,\tau) \Big)'' \right\}^2 dt
\end{equation}
where $\lam$ is a smoothing parameter that allows flexibility in how much a function is penalized for over-compensation. Accordingly, the problem of choosing the optimal value of $\lambda$ for the problem is commonly solved by generalized cross-validation (GCV) developed by \cite{CravenWahba}. For information on how $\hat{\bs{\beta}}_i$ is computed using \eqref{ls-penalized}, see \citet[p.~86-90]{RamsaySilverman2005}. 

Once the aging curves have been fitted, many direct anolgues for descriptive statistics of single numbers or vectors exist for functional data. For example, one can get an estimated mean aging curve by
\begin{equation}
\hat{\mu}(t) = \frac{\sum_{i=1}^N f_i(t)}{N}
\end{equation}
and an estimated variance function by 
\begin{equation}
\widehat{\text{Var}}(f_i(t))= \frac{\sum_{i=1}^N \big(f_i(t)-\hat{\mu}(t)\big)^2}{N-1}
\end{equation}

Furthermore, since B-spline expansion allows us to uniquely express each aging curve $i$ in our sample as a $J$-dimensional weight vector $\bs{\beta}_i$, any number of familiar multivariate analysis techniques are immediately made available to the problem of analysis of curves like the F-test, $k$-nearest neighbor clustering ($kNN$), and prinicpal components analysis (PCA). This is powerful because we have transferred the problem of analyzing a set of continuous curves with infinite points to one of analyzing a set of finite-dimensional vectors in $\mathbb{R}^J$.

One powerful technique in analysis of multivariate data is principal components analysis (PCA). For data that are vectors in $\mathbb{R}^J$, PCA is a method for defining a set of orthogonal basis vectors in $\mathbb{R}^J$ that most naturally expresses the variation in the data. Analogously for functional data, functional principal componenets analysis (fPCA) defines a set of orthogonal basis \textit{curves}, $\psi_1(t), \psi_2(t),\psi_3(t),\ldots$, that most naturally expresses the variation in the ensemble of curves. The first basis curve $\psi_1(t)$ explains the most variation in the data, the second basis curve $\psi_2(t)$ explains the most variation in the data given the first basis curve, and so on. 

In the multivariate case, the principal component scores are a useful way to re-express the original data in terms of the PCA basis. The PC score for the $i^{\textrm{th}}$ data point corresponding to the $j^{\textrm{th}}$ PC basis vector is a number that quantifies the extent to which the $i^{\textrm{th}}$ data point lies along the $j^{\textrm{th}}$ PC direction. Similarly in the functional case, we can speak of the PC score for the $i^{\textrm{th}}$ aging curve corresponding to the $j^{\textrm{th}}$ PC basis curve, denoted  $\xi_{ij}$. In essence, $\xi_{ij}$ is an aggregate measure of whether the $i^{\textrm{th}}$ aging curve tends to move in the same direction as $\psi_j(t)$ (i.e. positive $\xi_{ij}$), in the opposite direction (i.e. negative $\xi_{ij}$), or in a way that is unrelated to $\psi_j(t)$ (i.e. $\xi_{ij}$ close to zero). More formally, we can write the $i^{\textrm{th}}$ aging curves as 
\begin{equation}\label{pc}
f_i(t) = \mu(t) + \sum_{j=1}^{\infty} \xi_{ij} \psi_j(t)
\end{equation}
where $\mu(t)$ is an unknown mean curve for the functions $f_i(t)$. 

It is easy to see the usefulness of these scores: if the first PC $\psi_1(t)$ captures a sufficiently high proportion of the variation in our sample of curves, then just knowing these scores $\xi_{i1}$, $i=1, \ldots, N$ and the shape of the $\psi_1$ will give us a reliable idea of the shapes of each of the curves without having to examine the $N$ curves themselves. For more information on fPCA see \citet[p.~147-186]{RamsaySilverman2005}. 

\subsection{PACE: a method to handle sparse and irregular data.}

Traditional FDA methods work well when performance measurements are at the same ages $t_{1}, \ldots, t_{n}$  for each player and when $n$ is large. However, when this is not the case several problems arise. When $n$ is small the data is said to be sparse. Sparsity is a problem because it makes fitting the curves $f_i(t)$ very hard or even impossible. When measurements are not at the same ages for each player, the data is said to be irregular. Irregularity is a problem because the knot sequence $\tau$ is not the same for each $i = 1, \ldots, N$ which makes the $\bs{\beta}_i$ incomparable across $i$. Therefore, one cannot perform multivariate analysis, such as clustering, on $\bs{\beta}_i$.

In professional sports leagues, the average career is quite short. The average NFL career is 3.5 years, the average MLB career is 5.6 years, the average NHL career is 5.5 years, and the average NBA career is 4.8 years (\citealt{car-lengths}). Additionally, athletes start and end their careers at very different ages and may sometimes miss a year due to injury or simply not being signed by a team. So, sports data is often sparse and irregular.

In \citeyear{YaoMullerWang2005}, \citeauthor*{YaoMullerWang2005} proposed a method called Principal Components Analysis through Conditional Expectation (PACE) for irregular and sparse data that is assumed to come from a smooth function (\citealt*{YaoMullerWang2005}). They also have developed a package in MATLAB which can be downloaded from \url{http://www.stat.ucdavis.edu/PACE/}. We begin by assuming that for each datum $i = 1, \ldots, N$ we observe discrete performance metrics at ages $t_{i1}, \ldots, t_{i n_i}$ that come from smooth aging curves $f_i(t)$ with same unknown mean $\mu(t)$ and same unknown covariance $G(s,t)\equiv\cov(f_i(s), f_i(t))$ across $i$. The aging curves are also assumed to be independent across $i$.  

Borrowing the notation of equation \ref{pc}, the model we consider is 
\begin{equation} \label{pace-model}
y_{ij} = f_i(t_{ij}) + \eps_{ij} = \mu(t_{ij}) + \sum_{k=1}^{\infty} \xi_{ik} \psi_k(t_{ij}) + \eps_{ij}
\end{equation}
where the PC scores $\xi_{ik}$ are uncorrelated random variables across $k$ with mean 0 and variance $\lam_j$ and where $\eps_{ij}$ are additional measurement errors which are i.i.d with mean 0 and variance $\sigma^2$, independent of $\xi_{ik}$. In other words, discrete measurements $y_{ij}$ are realized values from a random aging curve that includes a measurement error. 

Replacing components of equation \ref{pace-model} by their estimates or predictions,  we have 
\begin{equation} \label{pace-f-hat}
\hat{f}_i(t) = \hat{\mu}(t) + \sum_{k=1}^{\infty} \hat{\xi}_{ik} \hat{\psi}_k(t)
\end{equation}
which holds for all $t$ in the range of ages in the entire sample, not just the range of ages of player $i$. That is, equation \ref{pace-f-hat} holds for all $t\in [\min{t_{i1}},\max{t_{in_{i}}}]$. We also highlight that in equation \ref{pace-f-hat},  the only componenet that depends on player $i$ are the predictions of player $i$'s PC scores, $\hat{\xi}_{ik}$. To predict the PC scores, we consider the conditional expectation given all the information about player $i$, namely, the measurements $y_i = (y_{i1}, \ldots, y_{i n_i})$. So we have
\begin{equation}\label{PCestimates}
\hat{\xi}_{ik} = \hat{\expec}(\xi_{ik} \cond y_i) = \hat{\lam}_k \hat{\psi}_{ik}(t) \hat{\Sigma}_{i}^{-1} (y_i - \hat{\mu}_i)
\end{equation}
where $\hat{\mu}_i = (\hat{\mu}(t_{i1}), \ldots, \hat{\mu}(t_{i n_i}))$, $\hat{\psi}_{ik}(t) = (\hat{\psi}_{k}(t_{i1}), \ldots, \hat{\psi}_{k}(t_{i n_i}))$, and $(\hat{\Sigma}_i)_{j,l} = \hat{G}(t_{i1}, t_{i n_i})$ $+ \hat{\sigma}^2 \ind{(j=1)}$. We omit discussion on how to get the estimates or predictions of the individual components that make up equation \ref{PCestimates}, but highlight that the quantities $\hat{\lam}_k$ and $\hat{\Sigma}_i$ are estimated from the entire data set. So $\hat{\xi}_{ik}$ borrows strength from the data on all players. For more information on how to get the estimates or predictions of the individual components of equation \ref{PCestimates}, see \cite{RiceSilverman1991} and \cite{CapraMuller1997}.

In practice, the prediction for $\hat{f}_i(t)$ uses the first $J$ eigenfunctions 
\begin{equation}
\hat{f}_i^J(t) = \hat{\mu}(t) + \sum_{k=1}^J \hat{\xi}_{ik} \hat{\psi}_k(t)
\end{equation}
\cite*{YaoMullerWang2005} suggest selecting $J$ through leave-one-out cross validation. Once one obtains estimates for the $J$ PC scores,  one can describe the aging curve with infinite points as a $J$-dimensional vector. So, one can perform their preferred multivariate clustering algorithm on these scores to get clusters of aging curves. 

\cite*{YaoMullerWang2005} show that \eqref{pace-f-hat} provides the best prediction assuming $\xi_{ik}$ and $\eps_{ij}$ are jointly Gaussian and works in the presence of sparsity. They also show that $\hat{\xi}_{ik}$ is the best linear prediction of $\xi_{ik}$ given the information from the $i^{th}$ player, irrespective of the Gaussian assumption. They also argue through simulation that the proposed method is robust to non-Gaussian error terms.
\section{Advantages of Functional Data Analysis in Sports}

An FDA approach offers several conceptual and assumptional advantages in studying aging curves in sports over conventional regression-based methods, in addition to allowing for a richer analysis.

Conceptually, a player's performance in reality changes continuously with age. Therefore treating the datum of a player's performance as a smooth curve rather than as a vector of discrete repeated measurements more closely resembles how we actually think about aging.

The functional approach also allows more flexibility in the assumptions of the data. FDA makes no assumptions on the dependence structure of repeated measurements of a player's performance which is needed in regression-based repeated measures analysis. The only assumption is that a player's performance changes as a smooth function with age; by doing so one implicitly takes into account the dependence of a player's performance across time and consequently does not have to specify the dependence structure of a given player's repeated measurements. Regression-based repeated measures analysis relies on accurately modelling the dependence structure of a given player's repeated measurements. If it is done poorly, results will be inaccurate. The FDA approach does not have this issue. 

Furthermore, in FDA there are no restrictions on the shape of the fitted functions and so we can obtain more nuanced mean curve shapes than we would with least squares regression with polynomial terms. One might obtain similarly nuanced shapes by additionally considering spline terms, but this involves a lot of trial and error.

FDA also allows for rich analysis because one can perform any number of familiar multivariate analysis techniques on the analysis of curves. For example, just as multivariate principal components analysis can tell us which linear combinations of variables most naturally explain the variation in data that are vectors, fPCA tells us which combinations of basis functions most naturally explain variation in our data set of curves (\citealt{RamsaySilverman2005}). Another example is that FDA allows for clustering of repeated measurements data.

PACE is particularly powerful in the study of aging curves as it offers two main advantages over the already advantageous traditional FDA methods. First, we are able to handle the problems of sparsity and irregularity that were discussed in section 2.2, as PACE was developed specifically for sparse and irregular data. Second, we are able to get fitted aging curves for each player across the entire range of ages in the sample, not just during the period of time they played. Essentially, this provides a tool for handling and imputing missing data. If we used ordinary regression with polynomial terms or even traditional FDA we would have to eliminate observations with missing data on the domain sequence. Thus, PACE has sample size advantages as well. 

\section{Hypothesis Testing of Aging Curves in MLB Players}

\subsection{Introduction}

Above, we have laid out a general argument for why FDA is a natural approach to take for analyzing sports data. In this section, we illuminate our argument with an application using real MLB offensive data and demonstrate the power of this functional approach by showing how it can be extended to perform hypothesis testing.

Specifically, we show that fPCA scores allow one to perform \textit{non-parametric} hypothesis testing on curves. Using this method, we test whether potential power hitters' and potential non-power hitters' (as identified by early-career power metrics) careers follow different trajectories. We find significant evidence that potential power hitters both peak earlier than potential non-power hitters and suffer a more drastic decline later in their careers.

This question of differential aging profiles is not new. In \citeyear{Fair2008}, \citeauthor{Fair2008} compared the overall aging trend of MLB players across decades.  In \citeyear{Sarris2012}, \citeauthor{Sarris2012} investigated this difference for players in the 2012 season. However, \citeauthor{Fair2008}'s \citeyear{Fair2008} analysis was based on a simplistic linear regression framework, while \citeauthor{Sarris2012}'s approach was not able to quantify the degree of uncertainty in the analysis. Using FDA, our approach transcends both of these difficulties.

\subsection{Data and Metrics}

To build metrics of production and obtain basic data on relevant players, we relied on the Lahman Baseball Database (LDB) and the Fangraphs.com's sabermetrics database.

For this paper, we consider only offensive production as measured by weighted on-base average (wOBA), a composite measure introduced in \citeyear{Tango2007} (\citealt{Tango2007}). wOBA is a linear function of offensive counts such as walks and home runs, but weighted with coefficients that fluctuate from year to year to account for differences in average pitching quality and other factors to make wOBA stable across different seasons. 

In order to minimize biases from missing data when performing smoothing, we restrict the sample to players who have a sufficiently contiguous time series of season statistics. Fortunately, since the LDB spans such a long period of time, we can afford to impose fairly strict requirements on the players' careers and still obtain a large enough sample for meaningful analysis.

First, we disregard all player-seasons with less than 200 plate appearances (PA's), since sabermetric research shows that the variability of many key basic batting statistics drops sharply after a player has seen 200 PA's. Secondly, we keep players with gaps in their MLB record of at most 1 season between the ages of 24 and 36. This filters out players who suffered from significant injuries or who spent long periods of time in the minor leagues between major-league stints, which may skew aging curves in a way that does not reflect natural aging.

We chose to sample players between the ages of 24 and 36 because existing studies suggest that players' batting ability peaks around age 28, so a window from age 24 to 36 gives us sufficient leeway in examining the aging profile around that peak (\citealt{Fair2008}; \citealt{Sarris2012}). Furthermore we allot more time for the post-peak period as the the phase where the players are in their decline is more interesting for practical purposes. Finally, we exclude observations prior to 1920 (the "Deadball Era"), when the rules of the game were drastically different and thus offensive statistics from that period are not comparable to statistics from the modern era.

After these restrictions, we are left with $N=217$ players. To demonstrate hypothesis testing, we also need to split the sample into players with high power potential and those without. To measure this, we use the player's average Isolated Slugging (ISO) in their age 24-25 seasons. ISO is calculated as slugging percentage minus batting average, and measures a players' ability to obtain extra-base hits while controlling for their overall batting average.

We define potential power hitters as those with age 24-25 average ISO $>$ 0.150 and potential non-power hitters as first-2 ISO $\leq$ 0.150. This is justified by data suggesting that league average ISO in the modern era is roughly 0.150. This threshold also conveniently splits our sample  of $n=217$ players into 104 potential non-power hitters and 113 potential power hitters.

\subsection{Fitting of Aging Curves and Main Modes of Variation}

To smooth these data for FDA, we use the penalized regression spline method described in section 2.1. We employ a cubic B-spline basis with knots at ages $(24, 26, 28, 30, 32, 34, 36)$ and use GCV to select the optimal value of the roughness penalty, $\lam=18.7$. In addition, we subtract each player's average wOBA across ages 24-36 from each measurement, i.e. for a fixed player $i$ and observation times $j=1, \ldots, n_i$ we considered measurements
\begin{equation}
\tilde{y}_{ij} = y_{ij} - \frac{\sum_{k=1}^{n_i} y_{ik}}{n_i}
\end{equation}
This was done because in this part of analysis, we are interested in how the \textit{modes of variation} in the aging curves differ across players. Without demeaning the data, most of the data's variation will likely be in magnitude of performance rather than in patterns of change of performance across time, within players. Call this demeaned measure the \textit{normalized wOBA}.

\begin{figure}[ht!]
\centering
\includegraphics[width=110mm]{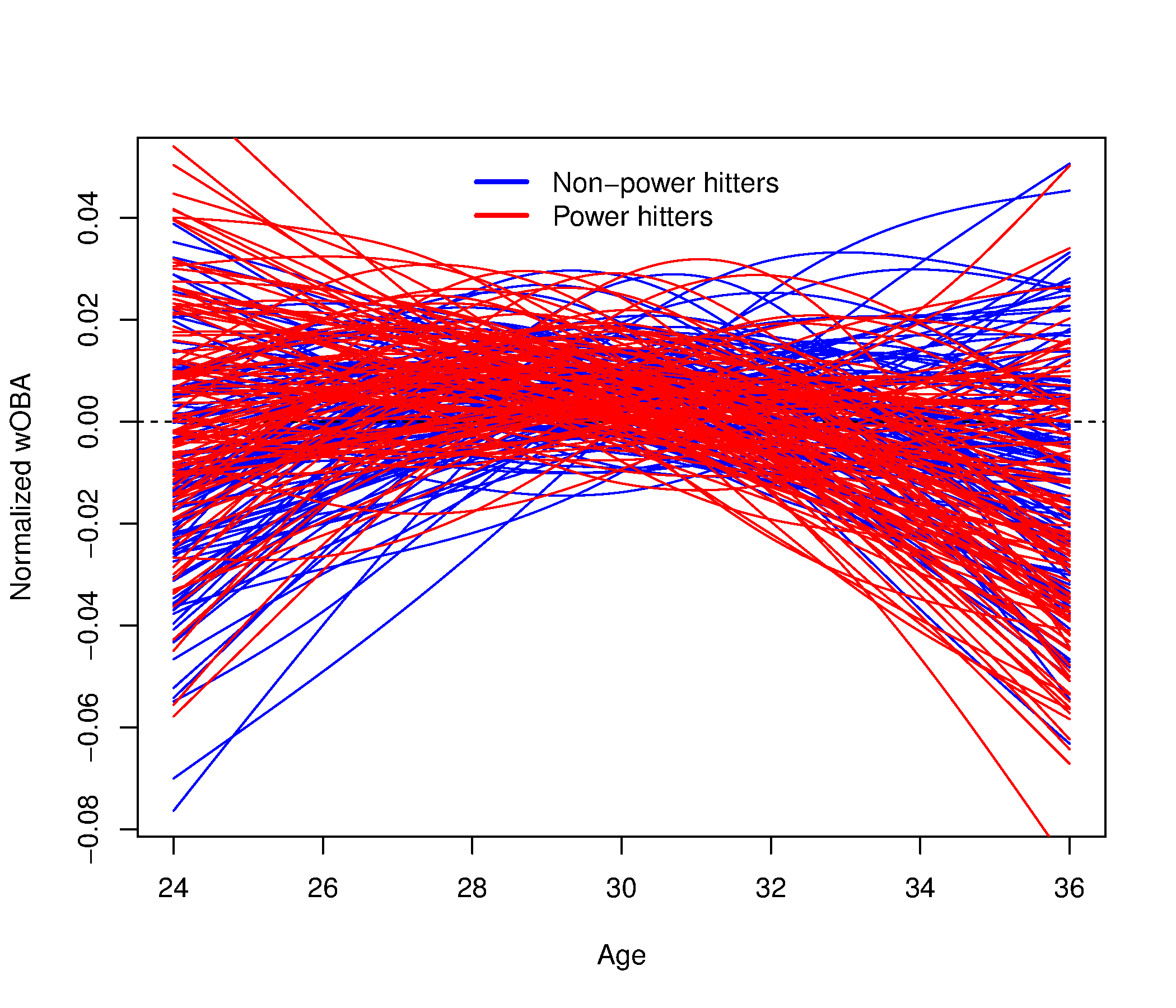}
\caption{Aging curves for players with high power potential and players with little power potential.}
\label{fig2}
\end{figure}

As evidenced by figure \ref{fig2}, there is an impressive variety of aging profiles in the data, not all of which mimic the classic upside-down ``U'' shaped curve. Complicating matters further, the large spectrum of shapes makes it difficult to discern from this figure whether or not there is any general difference between the aging profiles of power and non-power players, let alone whether such a difference is significant. Ideally, we would be able to focus our attention on only a couple major modes of variation. Since fPCA accomplishes exactly this, it is a key first step towards understanding our data and performing hypothesis testing.

\begin{figure}[H]
\centering
\begin{subfigure}{.5\textwidth}
  \centering
  \includegraphics[width=62mm]{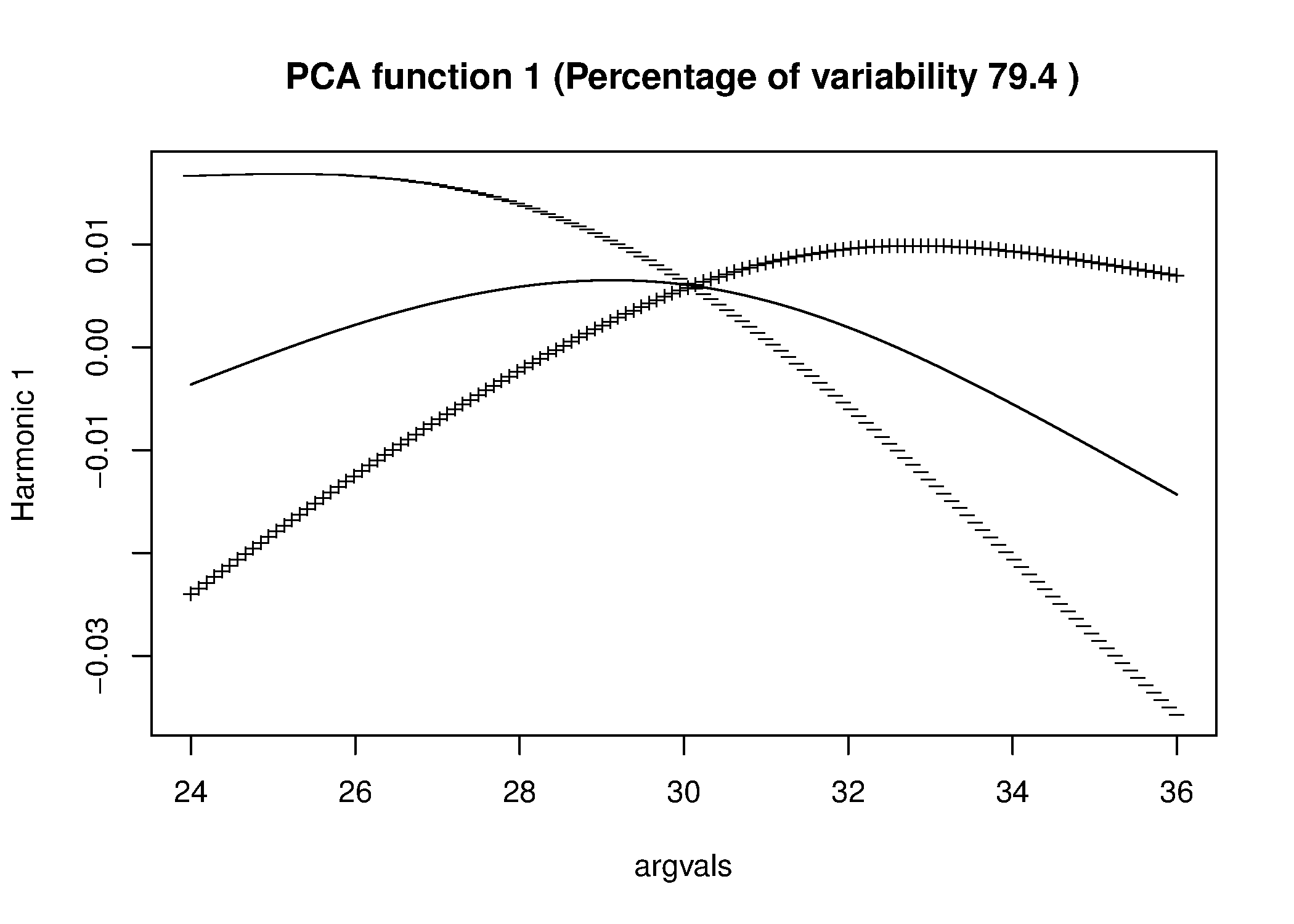}
  \caption{First principal component.}
  \label{fig3:sub1}
\end{subfigure}%
\begin{subfigure}{.5\textwidth}
  \centering
  \includegraphics[width=62mm]{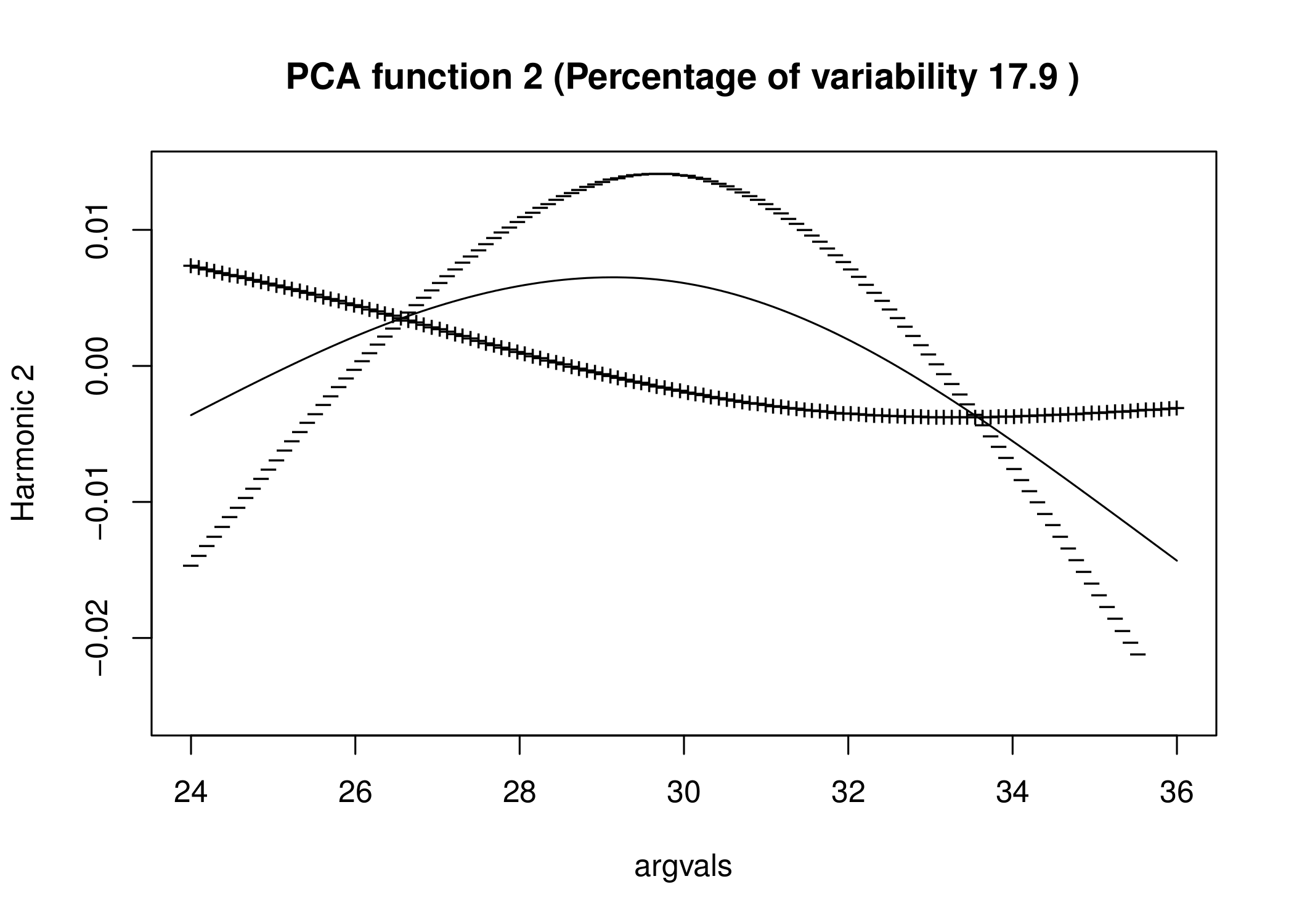}
  \caption{Second principal component.}
  \label{fig3:sub2}
\end{subfigure}
\caption{Plots of the first two principal components with aging tendencies for players with large negative and large positive scores.}
\label{fig3}
\end{figure}

We perform fPCA on our smoothed data and consider the first two PC functions, which together explain 97.3$\%$ of the variation in the data. The PC functions are plotted in Figure \ref{fig3} as deviations from the mean: the solid line is the mean aging tendency for all players in the sample, and the curve plotted as a series of ``$-$'' represents the aging tendency for a player with large negative scores corresponding to that PC and vice versa for the curve plotted as a series of ``$+$''. From the plots, we see that the main mode of variation is in a combination of rate of decline past their peak performance and whether their peak begins early or later. Players with high scores on PC1 tend to peak later and decline less rapidly, while players with low scores peak earlier and suffer a more drastic decline in later years. Similarly, the variation along the second principal component captures the ``peakiness'' of the players' careers, or in essence how consistent the players are over their careers.

\subsection{Hypothesis Testing: A Permutation Test Approach}

\begin{figure}[H]
\centering
\includegraphics[width=100mm]{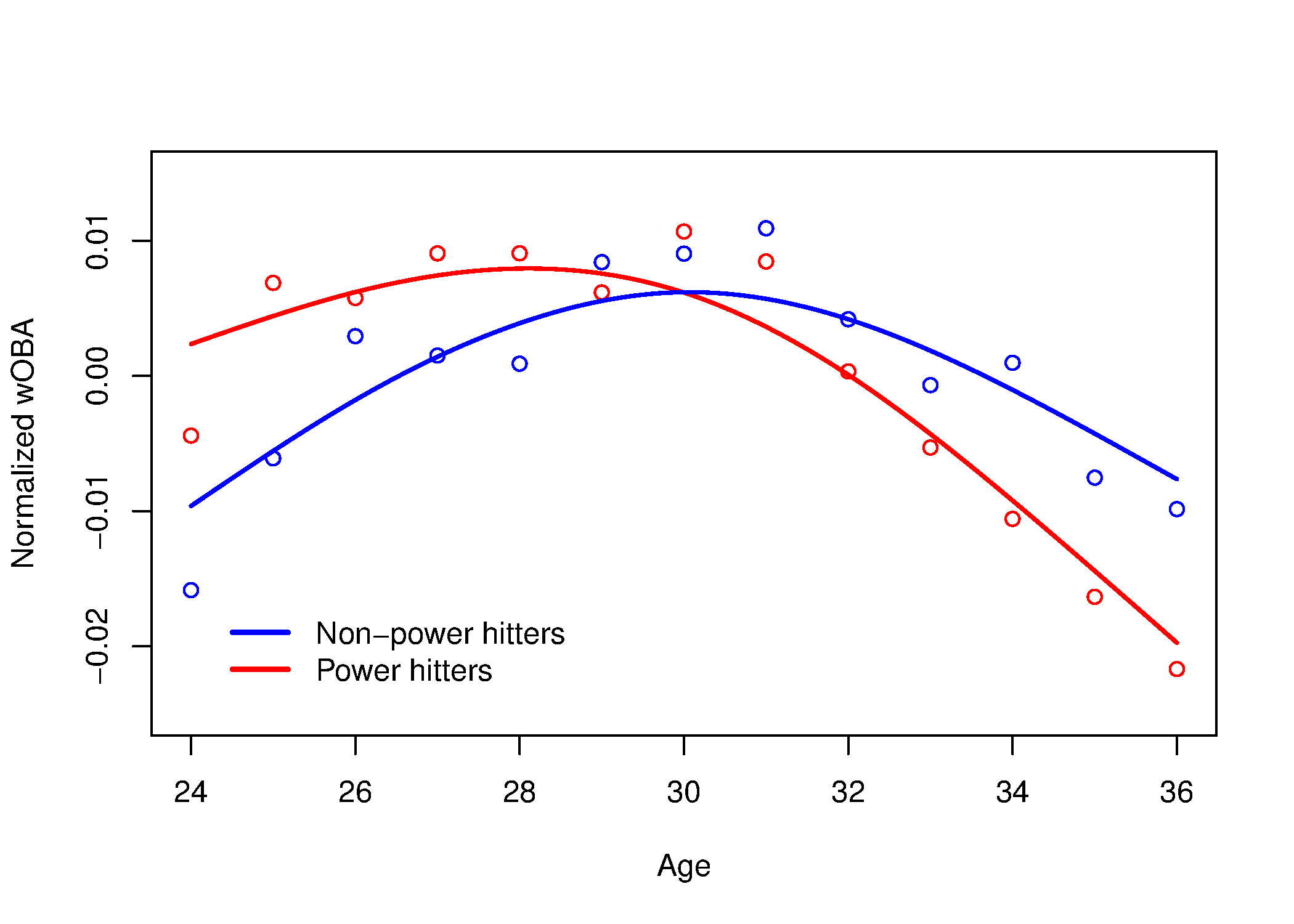}
\caption{Mean aging curves for players with high power potential and low power potential.}
\label{fig4}
\end{figure}

To test a hypothesis of the form ``$H_0$: no difference in aging curves between hitters with power potential and those without $H_a$: there is some difference,'' it is necessary to quantify the variation in curves for the power and non-power groups. As a first step, we calculate the \textit{mean curves} representing the general aging trends for each group, shown in figure \ref{fig4}.

On a casual visual inspection, it appears that the two groups have radically different aging profiles--potential power hitters both peak earlier and have a more severe decline in their later years. But we would like to test this rigorously.

In order to build a rigorous framework for hypothesis testing, one normally makes distributional assumptions on the data. The most natural assumption is that the data are somehow normally distributed; for example that players' performance at a fixed age is normally distributed. Then one may construct pointwise confidence intervals (\citealt[p.~70-73]{RamsaySilverman2005}). However, in practice, professional sports data is often not normal (\citealt{James1988}). This is particularly a problem when we consider players at later ages since players who have greater longevity are also players that are more talented. 

Our solution to non-normality is to consider a simple two-step non-parametric strategy which also considers an entire curve for a player as one unit, thereby implicitly taking into account the smooth correlation structure across time. Our procedure is as follows. First, we use fPCA to isolate the main mode of variation on the first principal component function (PC1) and extract the PC1 scores for each curve summarizing its variation in that direction. Then, we conduct a permutation test on the PC1 scores of each curve and simulate to obtain a $p$-value based on the empirical CDF. 

As shown in figure \ref{fig3}, PC1 explains $\approx$ 80\% of the variation in our sample, so we do not lose too much power in this assumption although the analysis is easy to extend to multiple PCs. The main idea is simple: if the power and non-power groups' aging profiles truly differ, then since the PC1 function explains the vast majority of the variance, then the two groups should have significantly different scores in that direction. Here, because the distribution of PC1 scores is not obvious, we propose a simulation-based permutation test of the PC scores, inspired by the DiProPerm methodology (Wei et al. \citeyear{wei2013}).

In general, suppose there are two groups in our data set, group 1 with $n$ observations and group 2 with $m$ observations (in our case, group 1 = power hitters and group 2 = non-power hitters). Denote the PC1 scores corresponding to group 1 by $P = (p_1, \ldots, p_n)$ and those corresponding to group 2 by $Q = (q_1, \ldots, q_m)$. Write $N = m + n$ and let the pooled sample be denoted
\begin{equation}
R = (r_1, \ldots, r_N) = (p_1, \ldots, p_n, q_1, \ldots, q_m)
\end{equation}
A natural sample statistic for the difference between scores is simply the absolute difference between the average score in group $P$ and group $Q$, denoted $T$:
\begin{equation}
T = \left| \frac{\sum_{i=1}^n r_i}{n} - \frac{\sum_{i=n+1}^N r_i}{m} \right| 
\end{equation}

\begin{figure}[H]
\centering
\includegraphics[width=100mm]{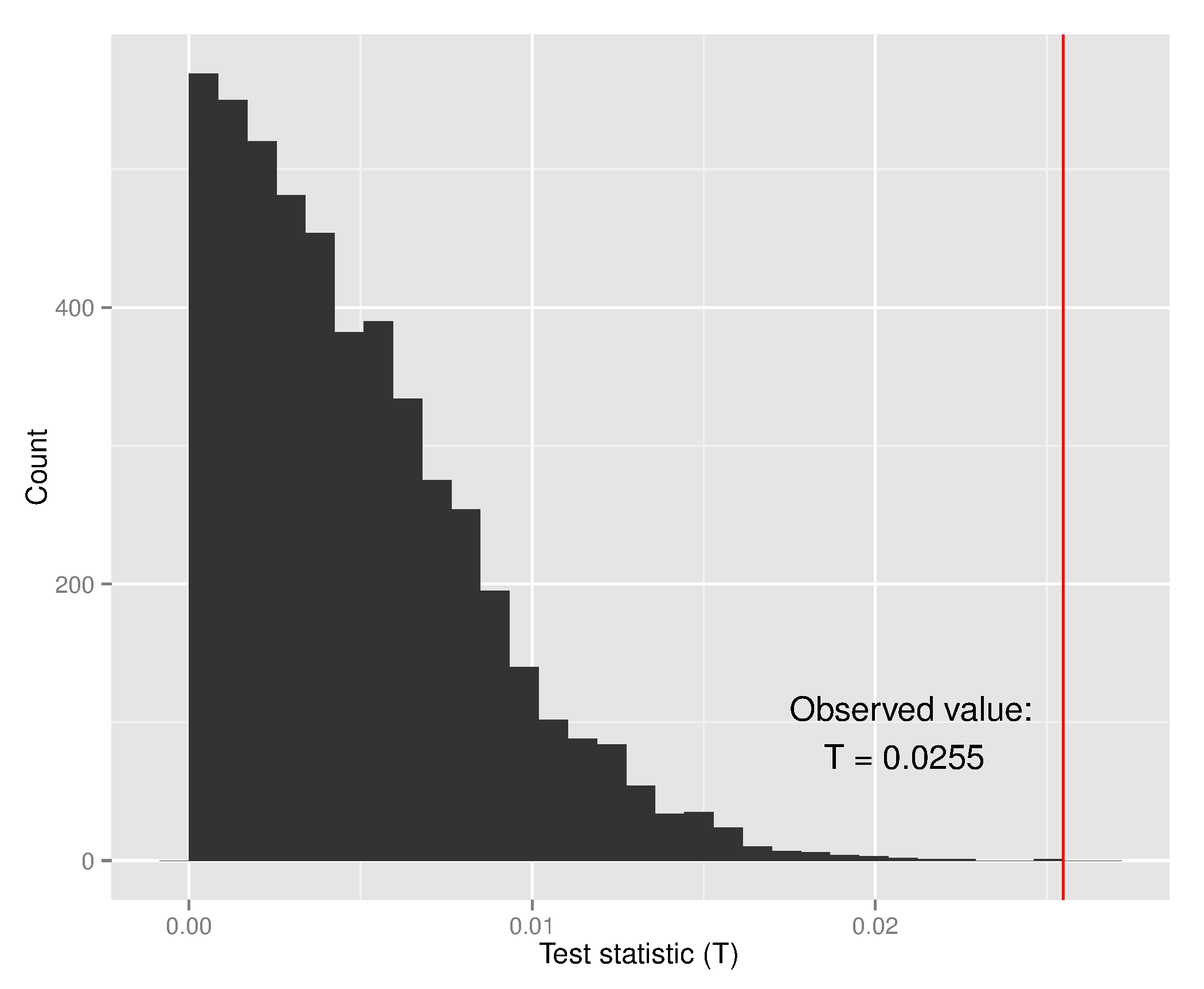}
\caption{Null distribution with observed sample statistic $T$ in red.}
\label{fig5}
\end{figure}

Now let $\pi = \{\pi(1), \ldots, \pi(N)\}$ denote a random permutation of $\{1, \ldots, N\}$. Then the sequence $R_{\pi} = (r_{\pi(1)}, \ldots, r_{\pi(N)})$ represents a random re-labeling of the subsets $P$, $Q$ of $R$ where the distinction between $P$ and $Q$ has been destroyed. Now recalculate our test statistic using this re-labelling:
\begin{equation}
T' = \left| \frac{\sum_{i=1}^n r_{\pi(i)}}{n} - \frac{\sum_{i=n+1}^N r_{\pi(i)}}{m} \right|
\end{equation}
Because $\pi$ is a random permutation, $T'$ can be viewed as the absolute difference in average PC1 score between the two groups under the null hypothesis of no difference in curves between the two groups. Therefore, if we repeatedly generate $\pi$ randomly and calculate $T'$, we obtain a \textit{null distribution} for $T$. Calculating the proportion of simulated $T'$ values that are greater than $T$ thus gives a reasonable $p$-value that can be objectively used to determine whether curves in two groups differ. Permutating and simulating on 5000 replications in this way, we obtain the null distribution in Figure \ref{fig5} with the observed sample statistic $T=0.0255$ indicated by the red line. 

This procedure allows us to reject the null hypothesis of no difference between power and non-power hitters with fairly high confidence, confirming our initial suspicions. Specifically, the proportion of simulated observations with $T' \geq 0.0255$ is $\approx 0$, so we can certainly reject at the 5\% significance level.

We conclude that the functional data approach to modeling data allows a uniquely flexible way to compare sets of curves, using a simple scalar function of the data (the PC1 score) that neatly summarizes the covariance structure inherent in the curves. Using this, we show that one can perform a hypothesis test for difference between curves with no distributional assumptions. As such, our method can be applied to performance metrics that are skewed or have otherwise unknown distributional properties as they often are in professional sports.

\section{Exploratory Analysis of Aging Curves in NBA Players}

\subsection{Introduction}

In this analysis, we illustrate the use of PACE described in section 2.2. Although using PACE on the analysis of the MLB data would allow us to increase sample size, we used traditional FDA for illustration purposes.  Since the MLB data set we worked with is much larger than the NBA data set, we were able to subset the data so that measurements were at the same times for each player and still have a reasonably large sample size. Sub-setting the NBA data in a similar way would result in either an unreasonably small sample size or functions fitted over an unreasonably small range of ages. 

Using PACE, we have two main goals. The first goal is to obtain summary statistics of the 645 NBA players in the sample, such as a mean curve and a variance curve. The second, and more interesting, goal is to find common patterns in how NBA player's performance changes with age and to find the types of players that follow which aging pattern. In satisfying this goal, we provide an example of how FDA allows for rich exploratory analysis. 

Although PACE will fit a continuous curve for each player across the entire domain sequence, not just during the period of time they played, the goal of this analysis did not include predicting aging curves of NBA players. If one wanted to predict aging curves of NBA players, one should use functional models that consider covariates that describe each NBA player. 

\subsection{Data and Metrics}

The data set, collected from \verb+http://www.basketball-reference.com/+, consists of all NBA players whose first season was the 1981-1982 season or later and played at least 8 seasons as of the 2012-2013 season. We only consider players with at least 8 seasons because we want to make sure we had enough seasons of experience to see aging patterns while not limiting sample size. We consider ages 19-39 which gave us at least 20 measurements for each age. This results in a sample size of 645 different players each with different repeated measurements.

The performance measure we consider is Win Shares originally developed by Bill James in \citeyear{James2002}. Win Shares is a popular one-number summary which estimates the number of wins a given player produces for his team in a given season by considering both the offensive and defensive performance of a given player. Since Win Shares is a per-season metric, rather than a per-age metric, we consider a player's age during a season to be their age on February 1st of that season. 

\subsection{Fitting of Aging Curves and Summary Statistics}

\begin{figure}[H]
\centering
\includegraphics[width=110mm]{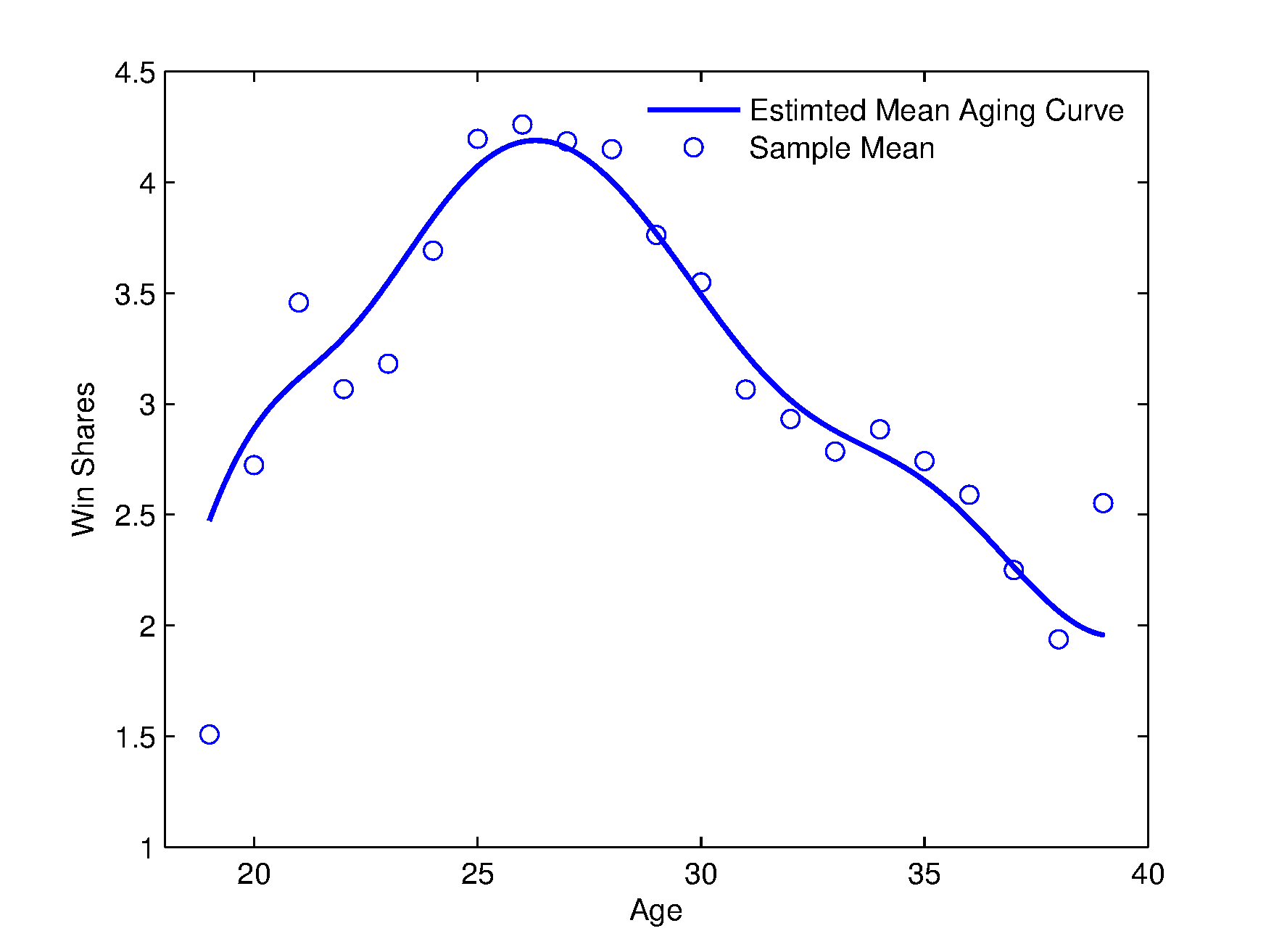}
\caption{Estimated mean curve for 645 NBA players.}
\label{fig6}
\end{figure}

\begin{figure}[H]
\centering
\includegraphics[width=110mm]{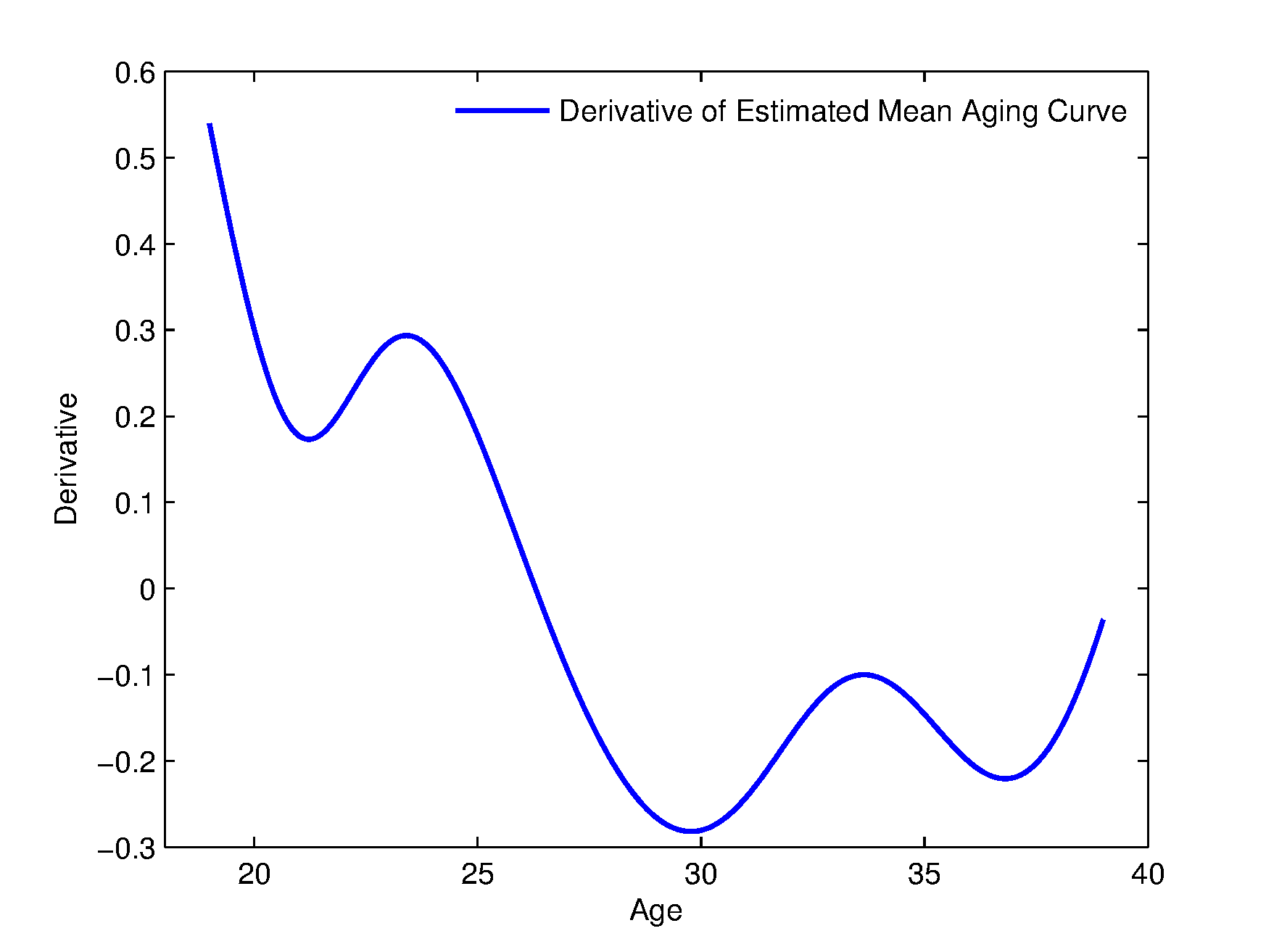}
\caption{First derivative of estimated mean curve.}
\label{fig7}
\end{figure}

After fitting aging curves $f_i(t)$ for each player $i$ in the sample using the methods described in section 2.2, we estimate the mean aging curve for all players across ages 19 through 39. The estimated mean curve $\hat{\mu}(t)$, is plotted in figure \ref{fig6} with the sample mean also plotted at each age. The mean curve shows peak performance at age 26.3147. The mean curve in conjunction with the first derivative of the mean curve in figure \ref{fig7} shows that NBA player's performance increases most rapidly between ages 19-20 and that there is another rapid increase between ages 23-25. Figures \ref{fig6} and \ref{fig7} also show that NBA player's performance decrease most rapidly at ages 29-31 and that there is another rapid decrease between ages 36-38.

\begin{figure}[H]
\centering
\includegraphics[width=110mm]{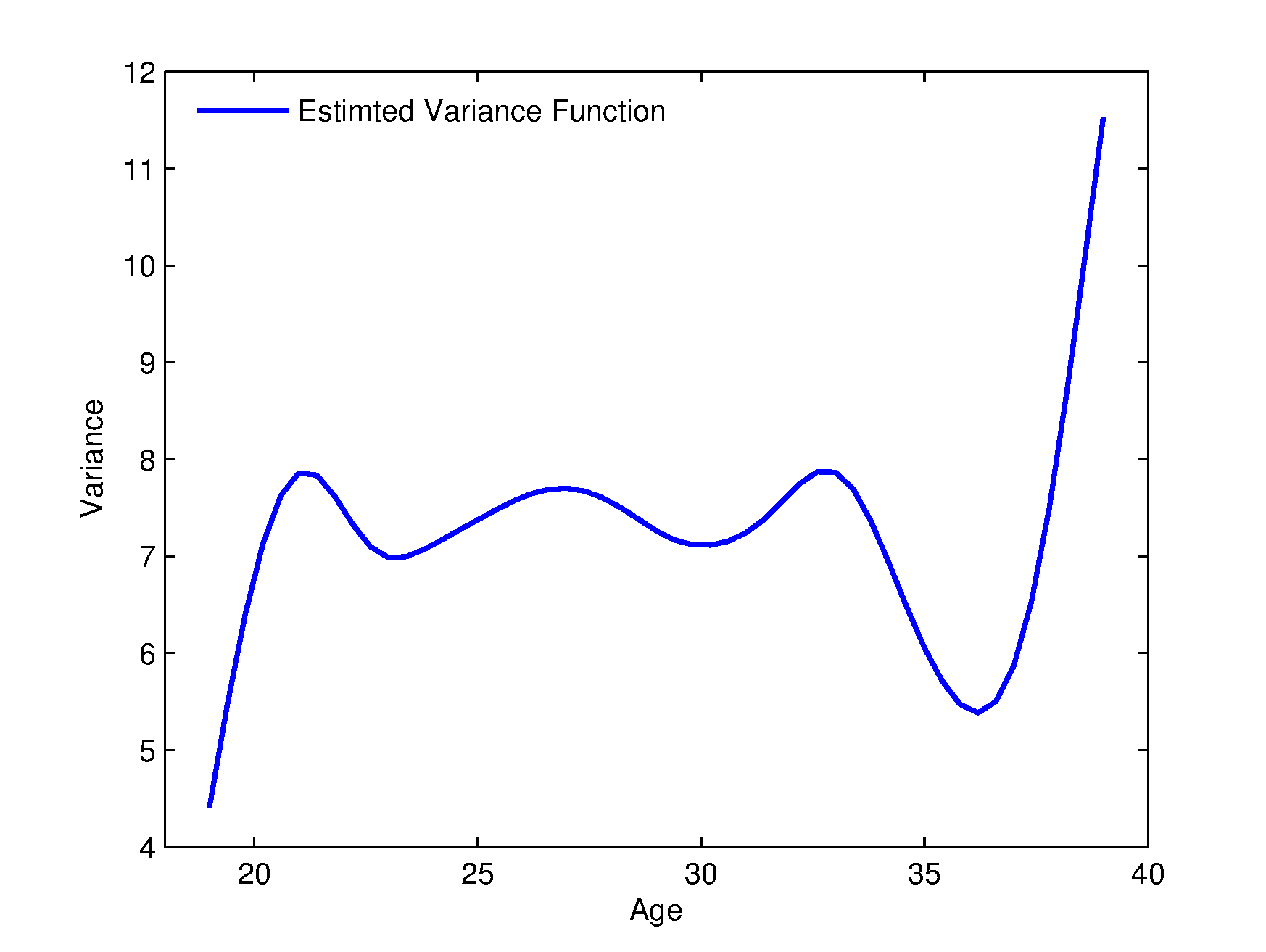}
\caption{Estimated variance function for 645 NBA players.}
\label{fig8}
\end{figure}

In the covariance function $G(s,t)$, if $s = t$ we may speak of a variance function $G(t,t)$. The variance function was also estimated using the methods described in section 5. The estimated variance function, $\hat{G}(t,t)$, is plotted in figure \ref{fig8}.

\subsection{Choosing Number of Clusters: how many distinct aging patterns, if any, are there?}

\begin{figure}[H]
\centering
\includegraphics[width=110mm]{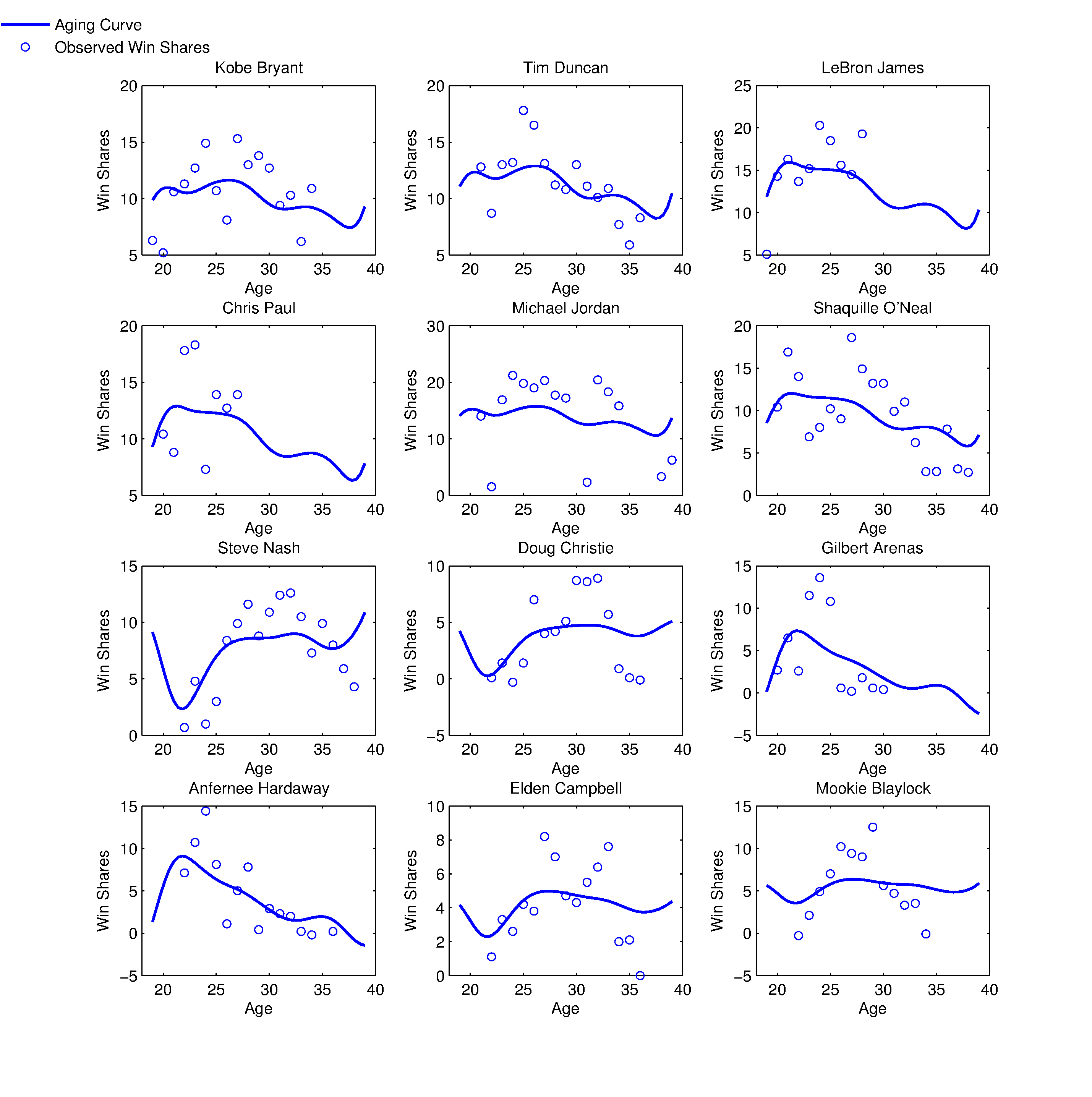}
\caption{Fitted curves for 12 example players.}
\label{fig9}
\end{figure}

Figure \ref{fig9} shows the aging curves for several example players. We consider current experienced high-level players in Kobe Bryant and Tim Duncan, young current high-level players in LeBron James and Chris Paul, experienced high-level players who have retired in Michael Jordan and Shaquille O'Neal, players who have aged in a very untraditional manner in Steve Nash and Doug Christie, players whose careers were cut short due to injury in Gilbert Arenas and Anfernee Hardaway, and experienced role players in Elden Campbell and Mookie Blaylock. Figure \ref{fig9} shows how insensitive the fitted aging curves are to off years in certain players. For example, note that the aging curve for Michael Jordan does not dramatically dip low to fit the low Win Shares values at ages 22 and 31 (Michael Jordan played only 18 games during the season he was 22 due to injury and only played 17 games during the season he was 31 due to coming out of retirement in the middle of the season). Also, note a few very common trends in these examples. The superstar players Kobe Bryant, Tim Duncan, Shaquille O'Neal, LeBron James, and Chris Paul all have very similar aging curves. These players very quickly reach a peak at age 20, keep this peak until around age 27, and then very slowly decline. We see that the role players Mookie Blaylock and Elden Campbell peak later than the superstar players in these examples. We see that the untraditional agers, Steve Nash and Doug Christie, have similarly shaped aging to each other and that the players with careers cut short, Gilbert Arenas and Anfernee Hardaway, also have similarly shaped aging curves. These common trends motivate clustering these curves to find common aging patterns and to find out what type of players fall into which cluster.

\begin{figure}[H]
\centering
\includegraphics[width=100mm]{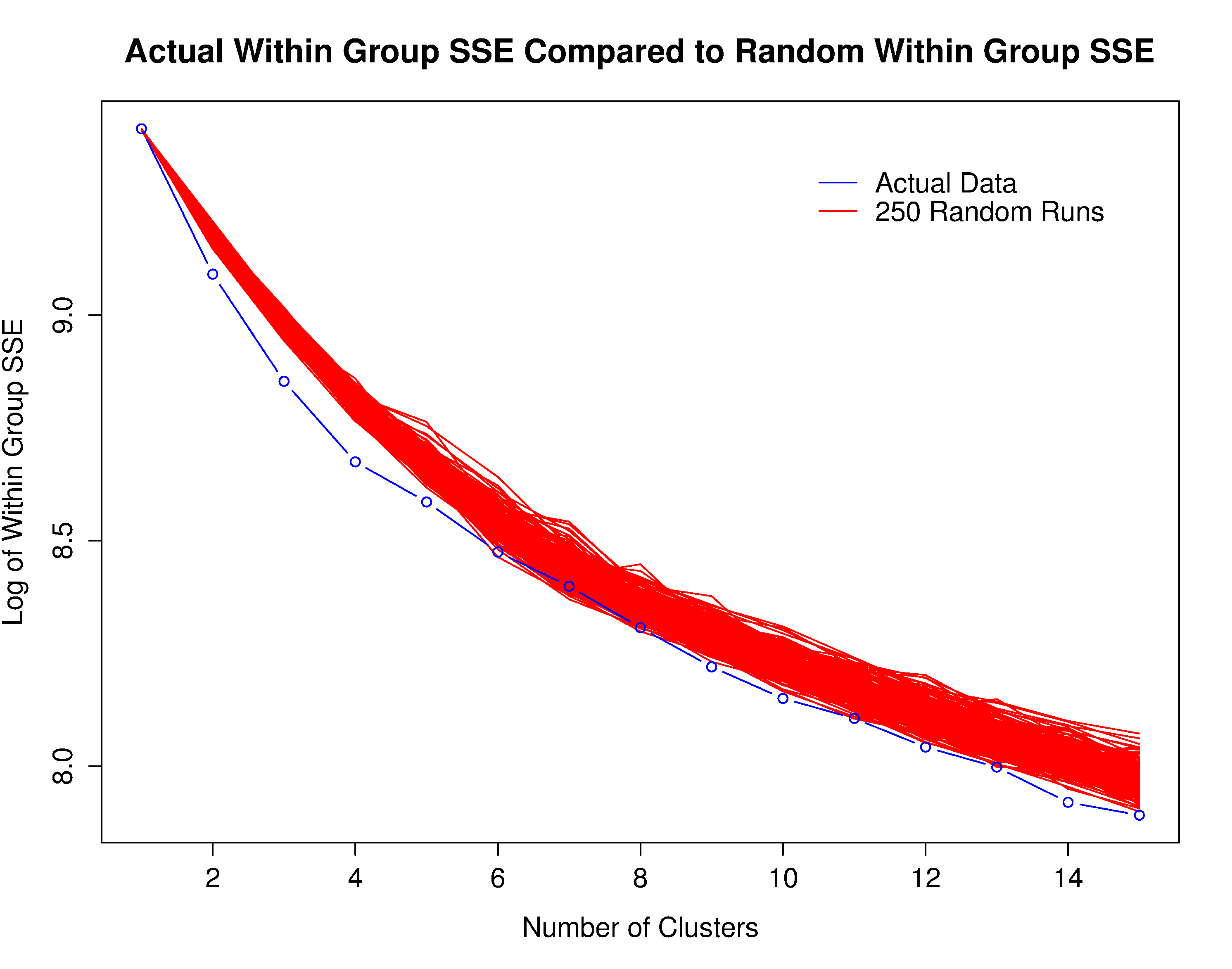}
\caption{Actual log of within group SSE compared to random runs.}
\label{fig10}
\end{figure}

\begin{figure}[H]
\centering
\begin{subfigure}{.5\textwidth}
  \centering
  \includegraphics[width=62mm]{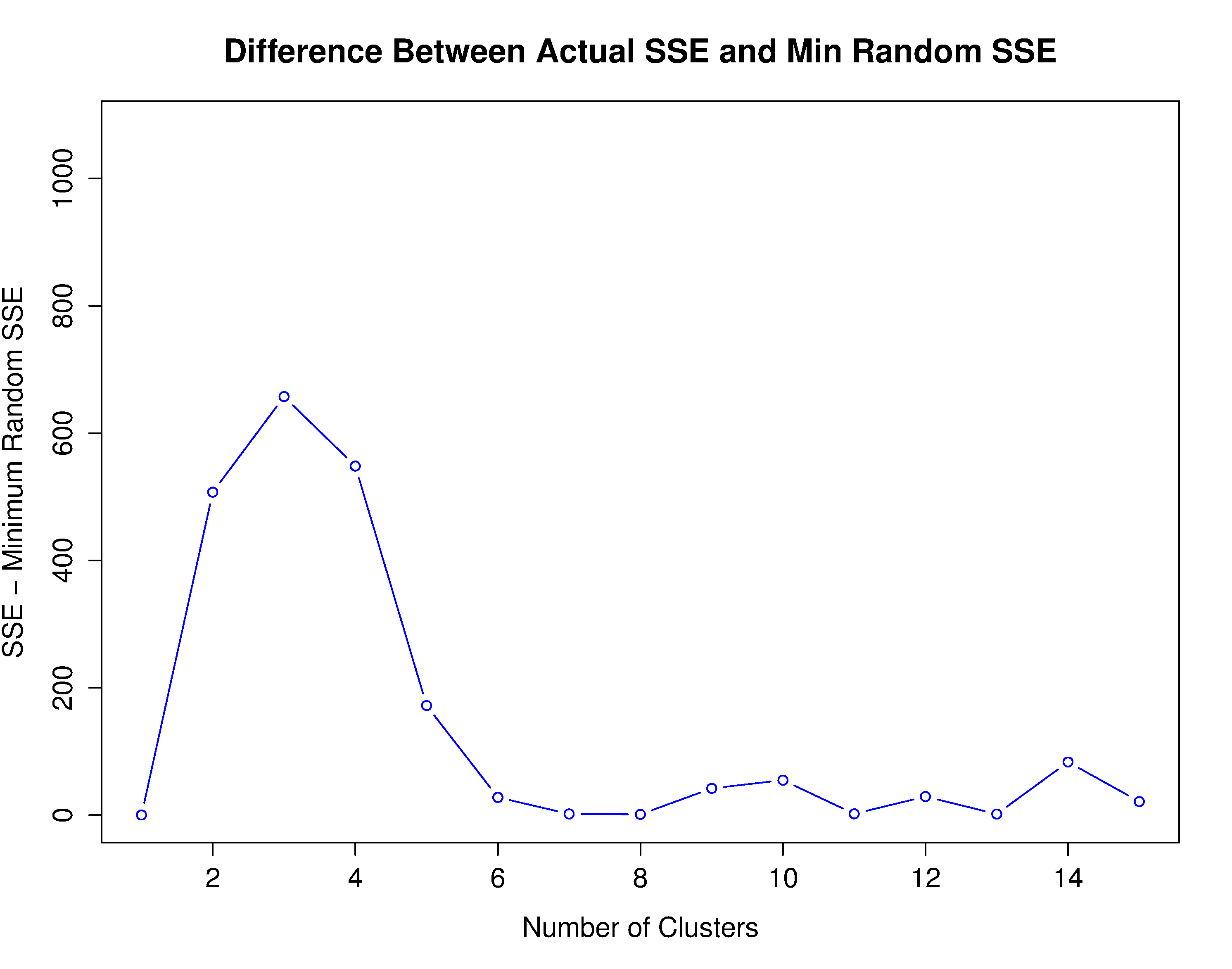}
  \caption{Compared to minimum of random runs.}
  \label{fig11:sub1}
\end{subfigure}%
\begin{subfigure}{.5\textwidth}
  \centering
  \includegraphics[width=62mm]{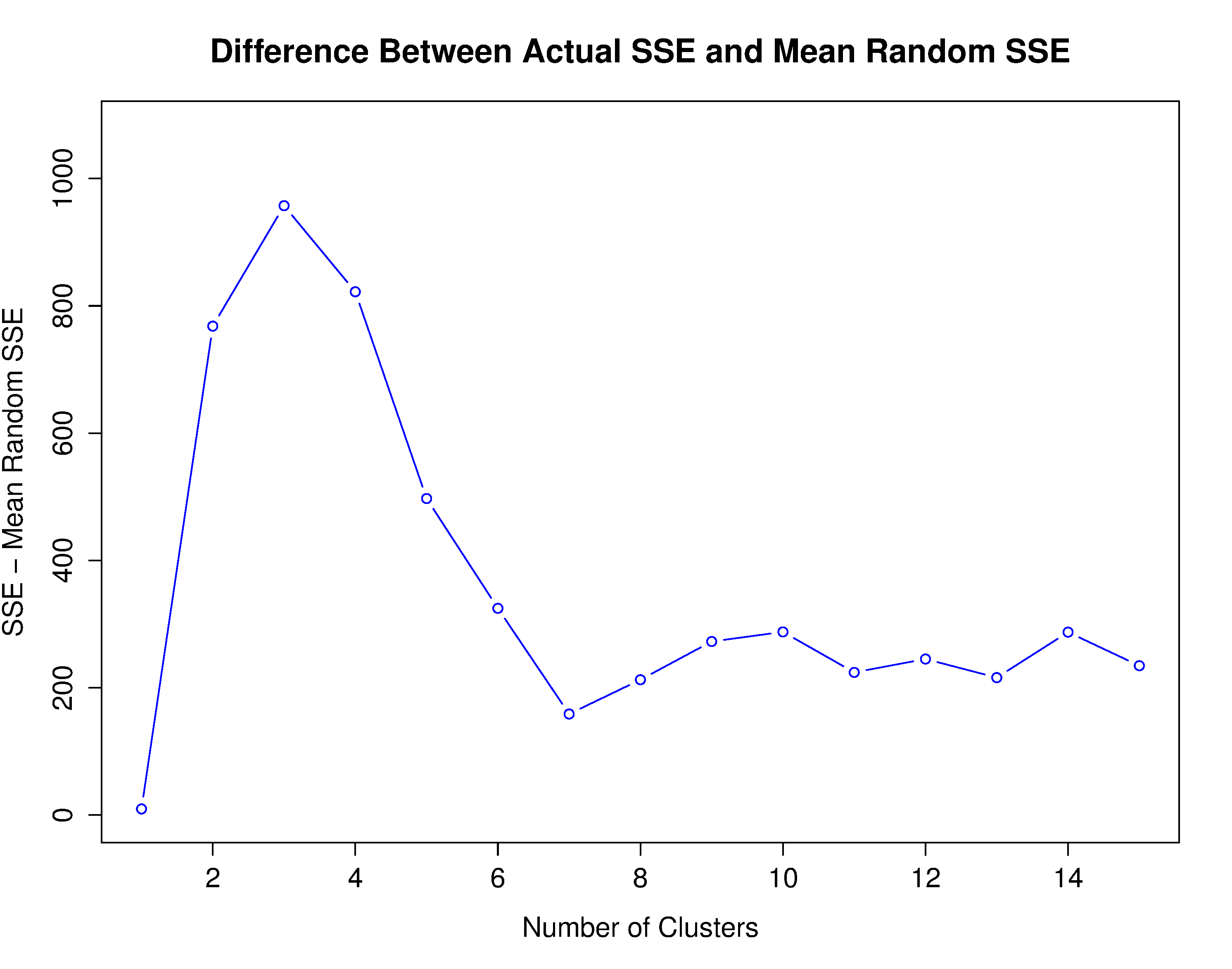}
  \caption{Compared to mean of random runs.}
  \label{fig11:sub2}
\end{subfigure}
\caption{Difference between actual within group SSE compared to random run.}
\label{fig11}
\end{figure}

For analysis with clustering, each player's performance is demeaned just as we did in the MLB analysis. We perform k-means clustering on the first 3 PC scores $\hat{\xi}_{ik}$. We select the first 3 principal components by cross-validation as described in section 2.2. We choose the number of clusters by considering using $k = 1, 2, \ldots, 15$ clusters. For each $k$, we calculate the actual within group sum of square error (SSE) and compare it to 250 random runs (see figure \ref{fig10}). In each random run, we randomly permute each of the 3 PC scores across the 645 players. This way, any inherent structure in the PC scores is removed from the data in the random permutations. Figure \ref{fig10} suggests that there truly is structure in the PC scores, particularly for $k = 2, 3, 4, 5$. In order to choose $k$, we compare the actual within group SSE to the minimum of the random within group SSEs (see figure \ref{fig11:sub1}) and compare the actual within group SSE to the mean of the random within group SSEs (see figure \ref{fig11:sub2}). Both of these suggest using $k = 3$ clusters. First, this analysis suggests that there are indeed distinct aging patterns as it would be reasonable to consider anywhere from 2 to 15 clusters. Second, this analysis suggests that the most reasonable number of clusters is 3. To see which of the three clusters the example players given earlier fall into, refer to table 4 in the appendix.

\subsection{Describing the Shape of the Three Clusters: what are the distinct aging patterns?}

\begin{figure}[H]
\centering
\includegraphics[width=108mm]{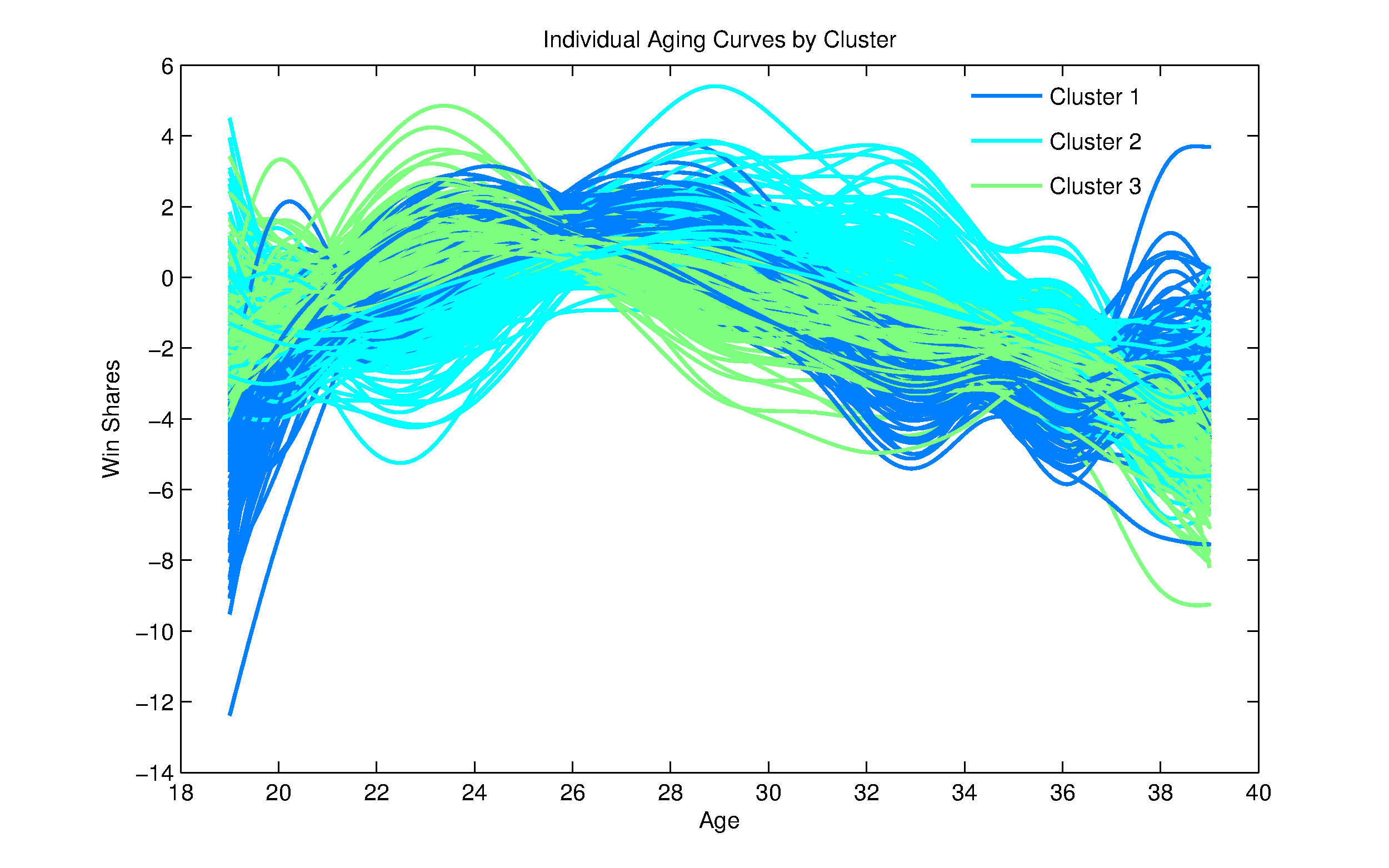}
\caption{Fitted aging curves of 645 players by cluster.}
\label{fig12}
\end{figure}

\begin{figure}[H]
\centering
\includegraphics[width=104mm]{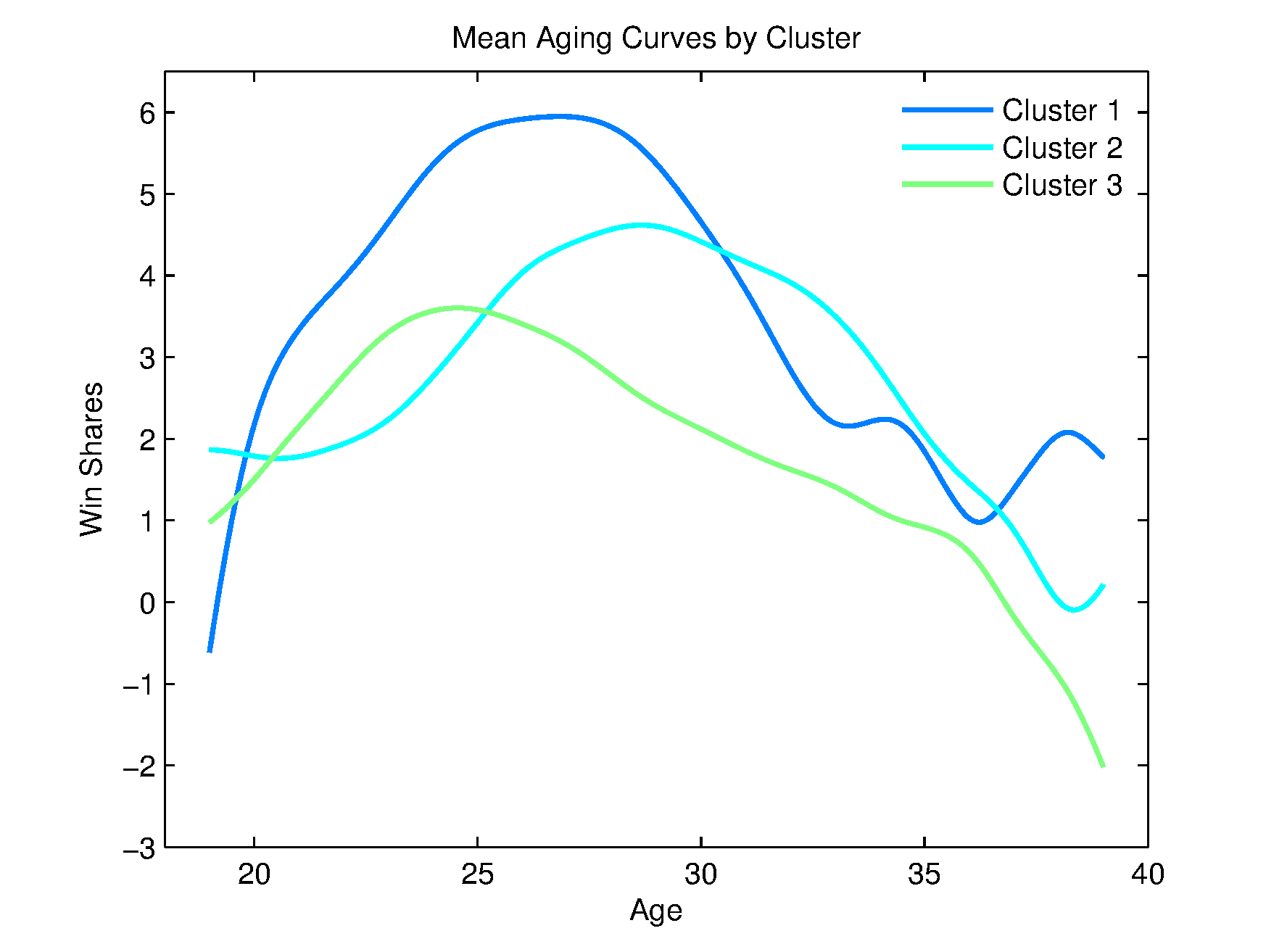}
\caption{Estimated mean curve for each cluster.}
\label{fig13}
\end{figure}

\begin{figure}[H]
\centering
\includegraphics[width=100mm]{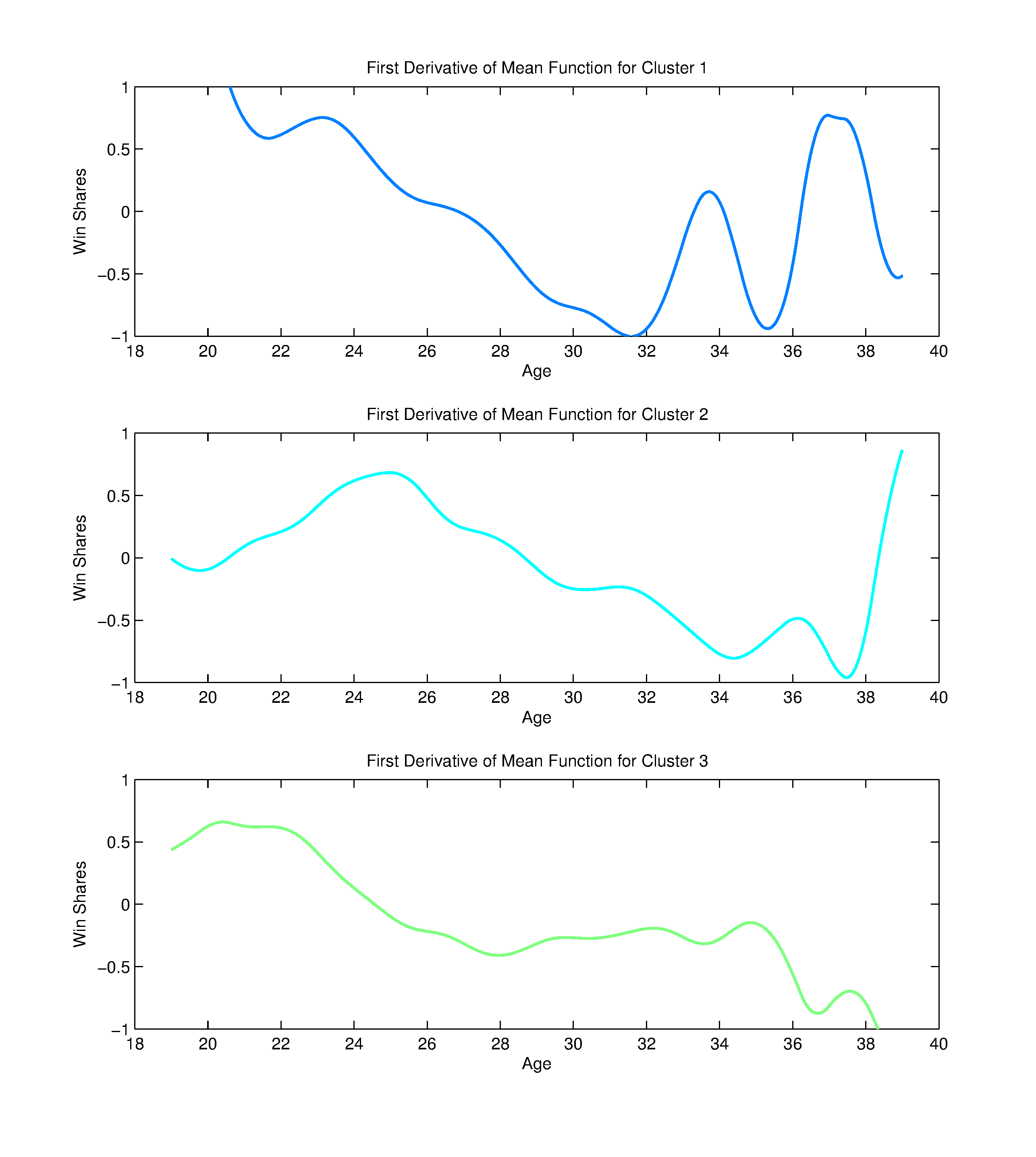}
\caption{Derivative of estimated mean curve for each cluster.}
\label{fig14}
\end{figure}

Having chosen 3 clusters, we begin to look into describing the shape of the mean aging curve of the three clusters. Figure \ref{fig12} shows each of the 645 demeaned fitted aging curves color coded by cluster. There are 159 players in cluster 1, 195 players in cluster 2, and 291 players in cluster 3.  In figure \ref{fig12}, one can clearly see shapes that are common to each cluster. We look at the mean curve for each cluster, as well as the derivatives, to help see the shapes in each cluster. Figure \ref{fig13} shows the mean curves for each cluster with individual player means $\sum_{k=1}^{n_i} y_{ik}/n_i$ added back in and figure \ref{fig14} shows the derivative of the mean curves for each cluster (y-axis different for each of the plots in figure \ref{fig14}).

Looking at figures \ref{fig13} and \ref{fig14}, we see that the aging pattern for cluster 1 peaks at age 26.84 with a mean Win Shares of 5.95.  This peak age is in between clusters 3 and 2 and so we call players in cluster 1 ``middle peakers''. The mean aging curve for cluster 1 enjoys the highest peak. Furthermore, if we define ``near-peak'' as the mean curve being within 10\% of the true peak, the mean curve for cluster 1 is at a near-peak level between the ages 23.80 and 29.03.  This is a longer near-peak than the other clusters.  They are also the top-performers from ages 19.82 to 30.44 and again from ages 36.64 to 39. We highlight specificity of peak age and age ranges that are possible when looking at a continous mean aging curve.

The aging pattern for cluster 2 peaks at age 28.67 with a mean Win Shares of 4.62. This is also the latest peak among the three clusters and so we call this cluster ``late peakers''. The mean curve for cluster 2 is at a near-peak level between the ages 26.28 and 31.04.  Cluster 2 are the top-performers from ages 30.44 to 36.64.

The aging pattern for cluster 3 peaks at age 24.56 with a mean Win Shares of 3.60. This is the earliest peak among the three clusters and so we call this cluster ``early peakers''. The mean aging curve for cluster 3 has the lowest peak. The mean curve for cluster 3 is at a near-peak level between the ages 22.86 and 26.69. This is the shortest near-peak among the clusters.  Furthermore, there is never a period of time where cluster 3 enjoys the highest performance. 

\subsection{Describing the Clusters: what type of players fall into which aging pattern?}

Now that we've described and plotted the shape of the mean aging curves in the three clusters we begin to look into the type of players that are in each cluster. First, we will consider the quality of performers in each cluster as measured by Win Shares. Second, we will begin to look more closely into differences between the abilities of the clusters by looking at differences in traditional box-score statistics, as well as True Shooting Percentage (TS\%) and Effective Field Goal \% (EFG\%), at the end of the player's career. Third, we will investigate the differences in player positions between the three clusters.	

We test differences in the quality of performers in each cluster in two ways. First, we test the differences in actual observed performance among the three clusters. We do this by comparing the mean observed Win Shares $\sum_{k=1}^{n_i} y_{ik}/n_i$ for each player among the three groups. Appealing to central limit theorem, we use a two-sided two-sample $t$-test to compare group differences. The mean observed Win Shares for each player in cluster 1 (``middle peakers'') is larger than the mean observed Win Shares for each player in cluster 2 (``late peakers'') (p-value $< 0.0001$) and the mean observed Win Shares for each player in cluster 2 is larger than the mean observed Win Shares for each player in cluster 3 (``early peakers'') (p-value $< 0.001$). Second, we test difference in performances across the entire age range 19-39 by comparing fitted aging curves along this range. We do this by calculating an integral of the fitted aging curves for each player among the three groups to get what we call an ``integral measure'' for each player. This integral measure calculates the area under the curve and allows us to put a value on the level of performance of a fitted curve across the entire age range 19-39. Again, we appeal to the central limit theorem and use a two-sided two-sample $t$-test to compare group differences. The mean integral measure for each player in cluster 1 (``middle peakers'') is larger than the mean integral measure for each player in cluster 2 (``late peakers'') (p-value of 0.0016) and the mean integral measure for each player in cluster 2 is larger than the mean integral measure for each player in cluster 3 (``early peakers'') (p-value $< 0.00001$). Based on these hypothes tests, it is safe to say that the ``middle peakers'' are the highest performers, the ``late peakers'' are the next best performers, and the ``early peakers'' are the worst performers.

We look into differences in several key statistics that describe a player. We consider end of career values for free throw percentage (FT\%), field goal percentage (FG\%),  three point percentage (3P\%), true shooting percentage (TS\%), effective field goal percentage (EFG\%), and per 36 minutes played (per 36) for blocks (BLK), assists (AST), free throws attempted (FTA), offensive rebounds (ORB), defensive rebounds (DRB), total rebounds (TRB), steals (STL), and points (PTS). Since we considered players at the end of their career, we only consider the 504 retired players in the data set. Appealing to central limit theorem, we use the appropriate one-sided two-sample $t$-test to compare group differences. Table 1 shows the means of the three clusters for the values above which have at least one significant difference between the three groups and table 2 shows the p-values. For the critical value for significance, we use the conservative Bonferroni correction on $\alpha = 0.05$ for the 39 $t$-tests to get a critical value of $\alpha = 0.05/39 = .001$.

\begin{table}[H]
\centering
\caption{Cluster Means}
\begin{tabular}{lllllll}
\hline
Variable & Cluster 1 & Cluster 2 & Cluster 3 &  \\
\hline
PTS per 36 &	15.39 & 13.40 & 13.66 \\
FT\% & 75.94	& 74.24 &	72.53 \\
3P\% & 28.39	& 26.04 &	23.69 \\
TS\% &	53.47 & 52.75 &	52.08 \\
\hline
\end{tabular}
\end{table}

\begin{table}[H]
\centering
\caption{P-Values}
\begin{tabular}{lllllll}
\hline
Variable & Clusters 1 vs. 2 & Clusters 1 vs. 3 & Clusters 2 vs. 3 \\
\hline
PTS per 36 & 3.21E-06	& 6.64E-06 &	0.479 \\
FT\% & 0.099	&2.17E-04 &	0.069 \\
3P\% & 0.095	&  4.02E-04 &	0.077 \\
TS\% & 0.055 &	1.52E-05 & 0.040 \\
\hline
\end{tabular}
\end{table}

Note that all the values which have at least one significant difference between the three groups are related to scoring. PTS per 36 measures scoring volume, TS\% measures scoring efficiency, and FT\% and 3P\%  are elements of scoring efficiency. Table 2 shows that cluster 1 scores significantly more points than both cluster 2 and cluster 3. Table 2 also shows that cluster 1 has a higher TS\%, FT\%, and 3P\% than cluster 3. This exploratory analysis suggests that scoring ability (both efficiency and volume) is correlated with aging pattern. Particularly, the ``middle peakers'' in cluster 1 tend to have a higher scoring ability. A possible reason is that scoring ability may tend to decrease faster with age than other skills. This exploratory analysis presented here motivates further analysis that looks at how scoring ability decreases with age compared with other basketball abilities. 

We also look at differences in positions between the three clusters by looking at the proportion of guards (G), wings (W), forwards (F), forward-centers (FC), and centers (C). We consider all 647 players in the sample. Table 3 shows the position proportions of the three clusters. We use the $\chi^2$-test of independence to compare the proportions between the three clusters and find that position and aging pattern are independent (p-value = 0.23). Note that this is contrary to the popular belief that different positions age differently (\citealt{Paine2009}). Using FDA to investigate whether aging pattern and position are independent, we are able to compare performance acros the entire range of ages 19-39 and not just the age of peak performance that other analyses have done  (\citealt{Paine2009}). This further illustrates the powerful analysis that can be done with FDA. We point the curious reader to figure \ref{fig15} in the appendix to see the mean aging curves for the demeaned data by position.

\begin{table}[!h]
\centering
\caption{Position proportions by cluster.}
\begin{tabular}{lllllll}
\hline
Position & Cluster 1 & Cluster 2 & Cluster 3 &  \\
\hline
G &	0.33333 & 0.32821 & 0.26804   \\
W & 0.16352 & 0.18974  & 0.16838 \\
F &	0.28931 & 0.18974 & 0.25430 \\
FC & 0.08805 & 0.14872 & 0.14089  \\
C &  0.16352 & 0.18974 & 0.16838  \\
\hline
\end{tabular}
\end{table}

\section{Conclusion}

This paper introduces the study of functional data analysis (FDA) to aging curves in sports. In particular, we recommend the PACE method develoepd by Yao, Muller, and Wang in \citeyear{YaoMullerWang2005} which was specifically developed for handling the sparse and irregular data that is often seen in an athlete's time series. We highlight the computational and assumptional advantages of an FDA approach as well as the richer analysis that is available. We then provide concrete examples by estimating an appropriately nuanced mean aging curve for both MLB and NBA players and by performing the rich analysis that we highlighed.

The analysis of MLB data demonstrates how the traditional FDA framework suggests a natural way to perform hypothesis testing for differences between aging curves in to way that can yield actionable conclusions. In particular, we show that the FDA approach offers a great advantage over traditional techniques when normality cannot be assumed, as is often the case in sports data. In addition our analysis demonstrates how fPCA allows one to summarize major modes of variation in aging curves, contributing intuition to our understanding of how different subsets of players age beyond the hypothesis testing setting. 

The analysis of aging curves in NBA players illustrates the use of the PACE method. We also illustrate the rich analysis that is possible in taking an FDA approach in studying aging curves in sports. In particular, we illustrate the exploratory clustering of aging curves that is possible and how it can motivate future analysis. This specific data analysis shows that there are three distinct aging patterns. Looking at the players that fall into each of the three patterns, we see that the clusters are very different in their overall ability as measured by Win Shares. More specifically, there are differences in scoring ability among the clusters which suggests that scoring ability is correlated with a player's aging pattern. This exploratory analysis motivates further analysis that looks at how scoring ability decreases with age compared with other basketball abilities. We also illustrate how the rich analysis of FDA can lead to conclusions that differ from analyses that use more rudimentary methods by showing that aging pattern is independent of position.

\section{Appendix}

\begin{table}[H]
\centering
\caption{Cluster of example players.}
\begin{tabular}{lllllll}
\hline
Player & Cluster \\
\hline
Kobe Bryant & 1   \\
Tim Duncan & 3 \\
LeBron James & 3 \\
Chris Paul & 3 \\
Michael Jordan & 3 \\
Shaquille O'Neal & 1 \\
Steve Nash & 2\\
Doug Christie & 2 \\
Gilbert Arenas & 3 \\
Anfernee Hardaway & 3 \\
Elden Campbell & 2\\
Mookie Blaylock & 1\\
\hline
\end{tabular}
\end{table}

\begin{figure}[H]
\centering
\includegraphics[width=100mm]{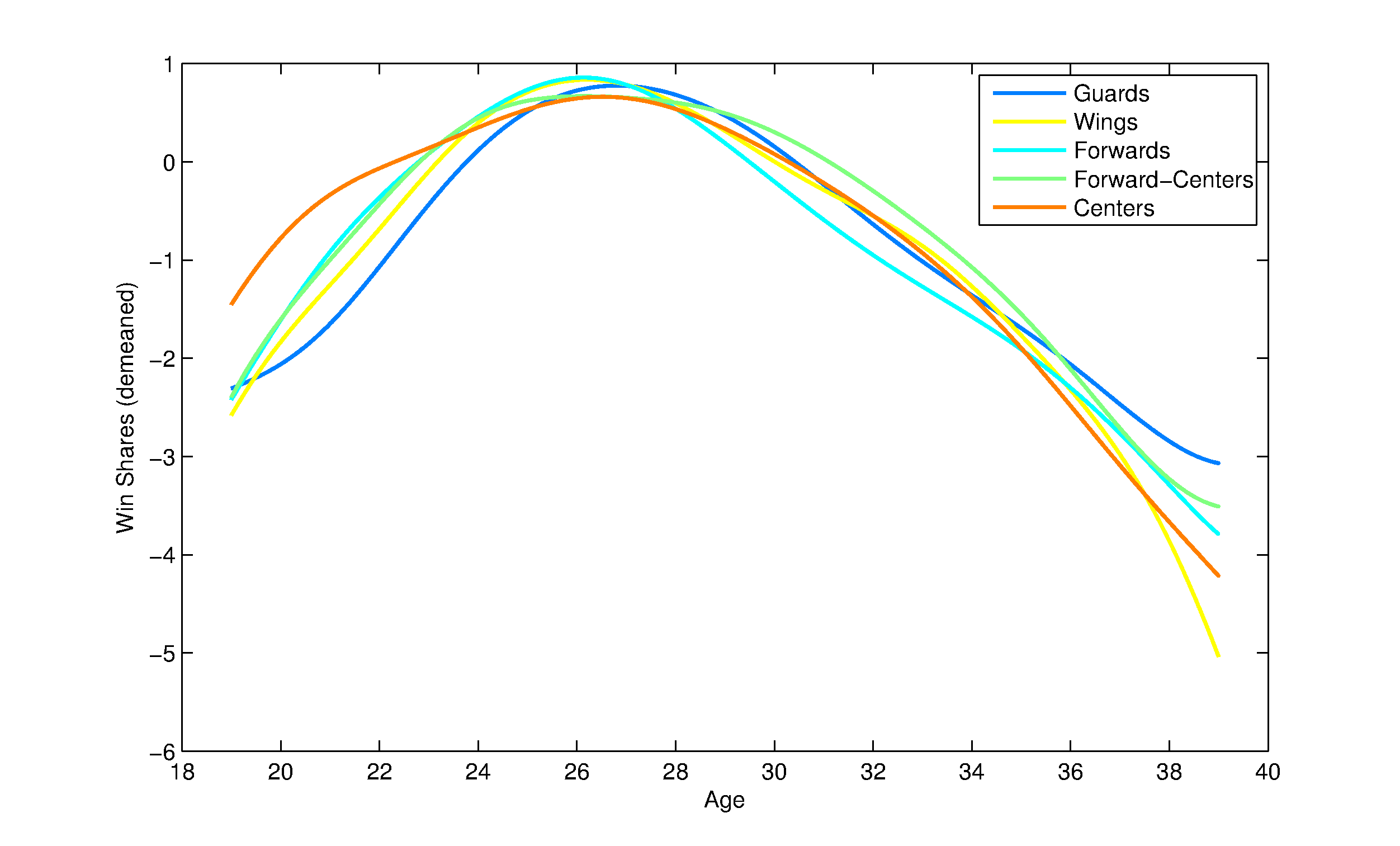}
\caption{Estimated mean curve for each position.}
\label{fig15}
\end{figure}

\bibliographystyle{chicago}
\bibliography{sample}

\end{document}